\documentclass{article}
\usepackage{spconf,amsmath,graphicx}
\usepackage{amssymb}
\usepackage[usenames]{color}
\title{Joint Maximum Sum-Rate Receiver Design and Power Adjustment for Multihop Wireless Sensor Networks \vspace{-0.0em}}
\twoauthors
  {Tong Wang, Rodrigo C. de Lamare}
    {Department of Electronics\\
    University of York, UK\\
    Email: tong.wang@hotmail.co.uk\\
~~~~~~~~~~rcdl500@ohm.york.ac.uk \vspace{-0.05em}}
  {Anke Schmeink}
    {UMIC Research Centre\\
    RWTH Aachen University\\
    D-52056 Aachen, Germany\\
Email: schmeink@umic.rwth-aachen.de \vspace{-0.05em}}
\begin{document}

\ninept

\maketitle
\begin{abstract}
In this paper, we consider a multihop wireless sensor network (WSN)
with multiple relay nodes for each hop where the amplify-and-forward
(AF) scheme is employed. We present a strategy to jointly design the
linear receiver and the power allocation parameters via an
alternating optimization approach that maximizes the sum rate of the
WSN. We derive constrained maximum sum-rate (MSR) expressions along
with an algorithm to compute the linear receiver and the power
allocation parameters with the optimal complex amplification
coefficients for each relay node. Computer simulations show good
performance of our proposed methods in terms of sum rate compared to
the method with equal power allocation.

\end{abstract}
\section{Introduction}

Recently, there has been a growing research interest in wireless
sensor networks (WSNs) as their unique features allow a wide
range of applications in the areas of defence, environment,
health and home \cite{Akyildiz}. They are usually composed of a
large number of densely deployed sensing devices which can
transmit their data to the desired user through multihop relays
\cite{Laneman}. Low complexity and high energy efficiency are the
most important design characteristics of communication protocols
\cite{Mitchell} and physical layer techniques employed for WSNs.
The performance and capacity of WSNs can be significantly enhanced
through exploitation of spatial diversity with cooperation between
the nodes \cite{Laneman}. In a cooperative WSN, nodes relay
signals to each other in order to propagate redundant copies of
the same signals to the destination nodes. Among the existing
relaying schemes, the amplify-and-forward (AF) and the
decode-and-forward (DF) are the most popular approaches
\cite{Hong}.

Due to limitations in sensor node power, computational capacity and
memory \cite{Akyildiz}, some power allocation methods have been
proposed for WSNs to obtain the best possible SNR or best possible
quality of service (QoS) \cite{Quek} at the destinations. The
majority of the previous literature considers a source and
destination pair, with one or more randomly placed relay nodes.
These relay nodes are usually placed with uniform distribution
\cite{Luo}, equal distance \cite{[4]}, or in line \cite{[10]} with
the source and destination. The reason of these simple
considerations is that they can simplify complex problems and obtain
closed-form solutions. A single relay AF system using mean channel
gain channel state information (CSI) is analyzed in \cite{[16]},
where the outage probability is the criterion used for optimization.
For DF systems, a near-optimal power allocation strategy called the
Fixed-Sum-Power with Equal-Ratio (FSP-ER) scheme based on partial
CSI has been developed in \cite{Luo}. This near-optimal scheme
allocates one half of the total power to the source node and splits
the remaining half equally among selected relay nodes. A node is
selected for relay if its mean channel gain to the destination is
above a threshold. Simulations show that this scheme significantly
outperforms existing power allocation schemes. One is the
'Constant-Power scheme' where all nodes serve as relay nodes and all
nodes including the source node and relay nodes transmit with the
same power. The other is the 'Best-Select scheme' where only the
node with the largest mean channel gain to the destination is chosen
as the relay node.

In this paper, we consider a general multihop wireless sensor
networks where the AF relaying scheme is employed. Our strategy is
to jointly design the linear maximum sum-rate (MSR) receiver
(\textbf{w}) and the power allocation parameter (\textbf{a}) that
contains the optimal complex amplification coefficients for each
relay node via an alternating optimization approach. It can be
considered as a constrained optimization problem where the objective
function is the sum-rate (SR) and the constraint is a bound on the
power levels among the relay nodes. Then the constrained MSR
solutions for the linear receiver and the power allocation parameter
can be derived. The proposed strategy and algorithm are not only
applicable to simple 2-hop WSNs but also to general multihop WSNs
with multi relay nodes and destination nodes. Another novelty is
that we make use of the Generalized Rayleigh Quotient \cite{Juday}
to solve the optimization problem in an alternating fashion.

This paper is organized as follows. Section 2 describes the multihop
WSN system model. Section 3 develops the joint MSR receiver design
and power allocation strategy. Section 4 presents the proposed
alternating optimization algorithm to maximize the sum rate. Section
5 presents and discusses the simulation results, while Section 6
provides some concluding remarks.

\section{System Model}
Consider a general m-hop wireless sensor network (WSN) with multiple
parallel relay nodes for each hop, as shown in Fig. 1.  The WSN
consists of $N_0$ source nodes, $N_m$ destination nodes and $N_r$
relay nodes which are separated into $m-1$ groups: $N_1$,$N_2$, ...
,$N_{m-1}$. We will focus on a time division scheme with perfect
synchronization, for which all signals are transmitted and received
in separate time slots. The sources first broadcast the $N_0\times1$
signal vector \textbf{s} to the first group of relay nodes. We
consider an amplify-and-forward (AF) cooperation protocol. Each
group of relay nodes receives the signal, amplifies and rebroadcasts
them to the next group of relay nodes (or the destination nodes). In
practice, we need to consider the constraints on the transmission
policy. For example, each transmitting node would transmit during
only one phase. In our WSN system, we assume that each group of
relay nodes transmits the signal to the nearest group of relay nodes
(or destination nodes)
directly. %We can use a block diagram to indicate the multihop
%WSN system as shown in Fig. 2.

\begin{figure}[!htb]
\begin{center}
\hspace*{0em}{\includegraphics[width=8cm, height=5cm]{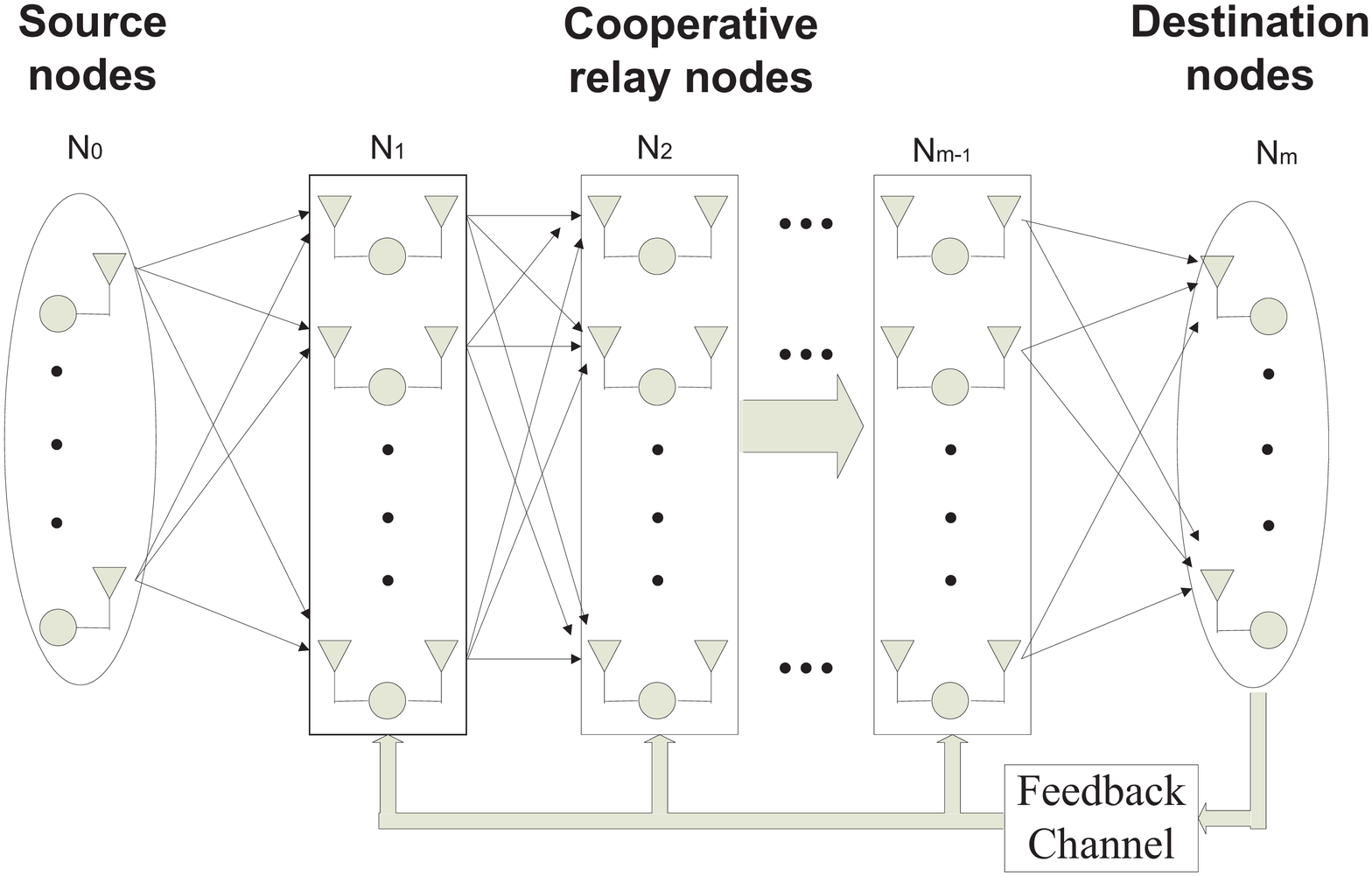}}
\vspace*{0.0em} \caption{\footnotesize $m$-hop WSN with $N_0$
sources, $N_m$ destinations and $N_r$ relays.}
\end{center}
\end{figure}
%\begin{figure}[!htb]
%\begin{center}
%\hspace*{0em}{\includegraphics[width=8.5cm, height=3cm]{Fig2_new.eps}}
%\vspace*{0.0em} \caption{Block diagram of the multihop WSN
%system.}
%\end{center}
%\end{figure}

Let $\textbf{H}_s$ denote the  $N_1\times{N_0}$
channel matrix between the source nodes and the first group of relay nodes,
$\textbf{H}_d$ denote the $N_m\times{N_{m-1}}$ channel
matrix between the $(m-1)$th group of relay nodes and destination nodes, and
$\textbf{H}_{i-1,i}$ denote the
$N_{i}\times{N_{i-1}}$ channel matrix between two groups of
relay nodes as described by
\begin{equation}
 \textbf{H}_s=
 \begin{bmatrix}
 \textbf{h}_{s,1}\\
 \textbf{h}_{s,2}\\
 \vdots\\
 \textbf{h}_{s,N_{1}}\\
 \end{bmatrix},~~~
 \textbf{H}_{d}=
 \begin{bmatrix}
 \textbf{h}_{m-1,1}\\
 \textbf{h}_{m-1,2}\\
 \vdots\\
 \textbf{h}_{m-1,N_{m}}\\
 \end{bmatrix},~~~
 \textbf{H}_{i-1,i}=
 \begin{bmatrix}
 \textbf{h}_{i-1,1}\\
 \textbf{h}_{i-1,2}\\
 \vdots\\
 \textbf{h}_{i-1,N_{i}}\\
 \end{bmatrix},~~~
\end{equation}
where $\textbf{h}_{s,j}=[h_{s,j,1},h_{s,j,2},...,h_{s,j,N_0}]$ for $j=1,2,...,N_1$ is a row vector between source nodes and the $j$th relay of the first group of relay nodes, $\textbf{h}_{m-1,j}=[h_{m-1,j,1},h_{m-1,j,2},...,h_{m-1,j,N_{m-1}}]$ for $j=1,2,...,N_m$ is a row vector between the $(m-1)$th group of relay nodes and the $j$th destination node and $\textbf{h}_{i-1,j}=[h_{i-1,j,1},h_{i-1,j,2},...,h_{i-1,j,N_{i-1}}]$ for $j=1,2,...,N_i$ is a row vector between the $(i-1)$th group of relay nodes and the $j$th relay of the $i$th group of relay nodes. The received signal at the $i$th group of relay nodes ($\textbf{x}_i$) for each phase can be expressed as:

Phase 1:

\begin{equation}
\textbf{x}_1=\textbf{H}_{s}\textbf{s}+\textbf{v}_{1},
\end{equation}
\begin{equation}
\textbf{y}_1=\textbf{F}_1\textbf{x}_1,
\end{equation}

Phase 2:
\begin{equation}
\textbf{x}_2=\textbf{H}_{1,2}\textbf{A}_1\textbf{y}_1+\textbf{v}_{2},
\end{equation}
\begin{equation}
\textbf{y}_2=\textbf{F}_2\textbf{x}_2,
\end{equation}

\hspace{8em} \vdots

Phase $i$: ($i=2, 3, ... , m-1$)
\begin{equation}
\textbf{x}_i=\textbf{H}_{i-1,i}\textbf{A}_{i-1}\textbf{y}_{i-1}+\textbf{v}_{i},
\end{equation}
\begin{equation}
\textbf{y}_i=\textbf{F}_i\textbf{x}_i,
\end{equation}

At the destination nodes, the received signal can be expressed as
\begin{equation}
\textbf{d}=\textbf{H}_{d}\textbf{A}_{m-1}\textbf{y}_{m-1}+\textbf{v}_d,
\end{equation}
where \textbf{v} is a zero-mean circularly symmetric complex
additive white Gaussian noise (AWGN) vector with covariance matrix
$\sigma^2\textbf{I}$. $\textbf{A}_i={\rm diag}\{a_{i,1},a_{i,2},...,a_{i,N_i}\}$ is a diagonal matrix whose
elements represent the amplification coefficient of each relay of
the $i$th group. $\textbf{F}_i$ denotes the normalization matrix which can normalize the power of the received signal for each relay of the $i$th group of relays.(see the appendix to find the expression of $\textbf{F}_i$.) Please note that the property of the matrix vector multiplication $\textbf{Ay}=\textbf{Ya}$ will be used in the next section, where \textbf{Y} is the diagonal matrix form of the vector \textbf{y} and \textbf{a} is the vector form of the diagonal matrix \textbf{A}. At the receiver, a linear MMSE detector is considered where the optimal filter and optimal amplification coefficients are calculated. The optimal amplification coefficients are transmitted to the relays through the feedback channel. And the block marked with a Q[$\cdot$] represents a decision device.

\section{Proposed Joint Maximum Sum-Rate Design of the Receiver and the Power Allocation}

By substituting (2)-(7) into (8), we can get
\begin{equation}
\begin{split}
\textbf{d} =& \textbf{C}_{0,m-1}\textbf{s} + \textbf{C}_{1,m-1}\textbf{v}_1 + \textbf{C}_{2,m-1}\textbf{v}_2\\
 &+ ...+\textbf{C}_{m-1,m-1}\textbf{v}_{m-1}+\textbf{v}_d\\
=&\textbf{C}_{0,m-1}s + \sum_{i=1}^{m-1}\textbf{C}_{i,m-1}\textbf{v}_i+\textbf{v}_d
\end{split}
\end{equation}
where
\begin{equation}
\textbf{C}_{i,j} = \left\{ \begin{array}{ll}
\prod_{k=i}^j\textbf{B}_k, & \textrm{if $i\leqslant j$,}\\\textbf{I}, & \textrm{if $i>j$}.\\
\end{array}\right.
\end{equation}
and
\begin{equation}
\textbf{B}_0 = \textbf{H}_s
\end{equation}
\begin{equation}
\textbf{B}_i = \textbf{H}_{i,i+1}\textbf{A}_i\textbf{F}_i~~~~~\textrm{for}~i=1,~2,~...,~m-2
\end{equation}
\begin{equation}
\textbf{B}_{m-1} = \textbf{H}_d\textbf{A}_{m-1}\textbf{y}_{m-1}
\end{equation}
We focus on the system which consists of one source node. Therefore, the expression of the Sum Rate (SR) for our m-hop WSNs is expressed as
\begin{equation}
{\rm SR}=\frac{1}{m}\log_2\left[1+\frac{\sigma_s^2}{\sigma_n^2}\frac{\textbf{w}^H\textbf{C}_{0,m-1}\textbf{C}_{0,m-1}^H\textbf{w}}{\textbf{w}^H(\sum_{i=1}^m\textbf{C}_{i,m-1}\textbf{C}_{i,m-1}^H)\textbf{w}}\right]
\end{equation}
where $\textbf{w}$ is the linear receiver, and $(\cdot)^H$ denotes the complex-conjugate (Hermitian) transpose. Let
\begin{equation}
\boldsymbol{\phi} = \textbf{C}_{0,m-1}\textbf{C}_{0,m-1}^H
\end{equation}
and
\begin{equation}
\textbf{Z} = \sum_{i=1}^m\textbf{C}_{i,m-1}\textbf{C}_{i,m-1}^H
\end{equation}
The expression for the sum-rate can be written as
\begin{equation}
{\rm SR}=\frac{1}{m}\log_2\left(1+\frac{\sigma_s^2}{\sigma_n^2}\frac{\textbf{w}^H\boldsymbol{\phi}\textbf{w}}{\textbf{w}^H\textbf{Z}\textbf{w}}\right)=\frac{1}{m}\log_2(1+ax)
\end{equation}
where
\begin{equation}
a =\frac{\sigma_s^2}{\sigma_n^2}
\end{equation}
and
\begin{equation}
x =\frac{\textbf{w}^H\boldsymbol{\phi}\textbf{w}}{\textbf{w}^H\textbf{Z}\textbf{w}}
\end{equation}
Since $\frac{1}{m}\log_2(1+ax)$ is a monotonically increasing function of $x$ ($a>0$), the problem of maximizing the sum rate is equivalent to maximizing $x$. In this section, we consider the case where the total power of the relay nodes in each group is limited to some value $P_{T,i}$ (local constraint). The proposed method can be considered as the following optimization problem:
\begin{equation}
\begin{split}
&[\textbf{w}_{opt},\textbf{a}_{1,opt},...,\textbf{a}_{m-1,opt}] ~=\arg\max_{\textbf{w},\textbf{a}_1,...,\textbf{a}_{m-1}}\frac{\textbf{w}^H\boldsymbol{\phi}\textbf{w}}{\textbf{w}^H\textbf{Z}\textbf{w}},  \\
&{~~~~~~~~~\textrm {subject to}}  ~ P_i=P_{T,i},~i=1,2,...,m-1.
\end{split}
\end{equation}
where $P_i$ as defined above is the transmitted power of the $i$th
group of relays, and $P_i=N_{i+1}\textbf{a}_i^H\textbf{a}_i$. We can
notice that the expression
$\frac{\textbf{w}^H\boldsymbol{\phi}\textbf{w}}{\textbf{w}^H\textbf{Z}\textbf{w}}$
in (20) is the Generalized Rayleigh Quotient, therefore the optimal
solution of our maximization problem can be solved:
$\textbf{w}_{opt}$ is any eigenvector corresponding to the dominant
eigenvalue of $\textbf{Z}^{-1}\boldsymbol{\phi}$.

In order to obtain the optimal power allocation vector
$\textbf{a}_{opt}$,  we rewrite
$\frac{\textbf{w}^H\boldsymbol{\phi}\textbf{w}}{\textbf{w}^H\textbf{Z}\textbf{w}}$
and the expression is given by
\begin{equation}
\frac{\textbf{w}^H{\boldsymbol \phi}\textbf{w}} {\textbf{w}^H
\textbf{Z} \textbf{w}} = \frac{\textbf{a}_i^H \textbf{M}_i
\textbf{a}_i} {\textbf{a}_i^H {\rm diag} \{\textbf{w}_i^H
\textbf{P}_i \textbf{w}_i\} \textbf{a}_i + \textbf{w}_i^H
\textbf{T}_i\textbf{w}_i}, ~{\rm for}~i=1,~2,...,~m-1,
\end{equation}
where\\ $\textbf{M}_i = {\rm
diag}\{\textbf{w}_i^H\textbf{C}_{i+1,m-1}\textbf{H}_{i,i+1}\textbf{F}_i\}\textbf{C}_{0,i-1}\textbf{C}_{0,i-1}^H \times$\\
${\rm
diag}\{\textbf{F}_i^H\textbf{H}_{i,i+1}^H\textbf{C}_{i+1,m-1}^H\textbf{w}_i\}$,
\\ \\ $\textbf{P}_i =
\textbf{C}_{i+1,m-1}\textbf{H}_{i,i+1}\textbf{F}_i\}(\sum_{k=1}^i\textbf{C}_{k,i-1}\textbf{C}_{k,i-1}^H)
\times $\\ ${\rm
diag}\{\textbf{F}_i^H\textbf{H}_{i,i+1}^H\textbf{C}_{i+1,m-1}^H \}$
\\  and \\ $\textbf{T}_i =
(\sum_{k=i+1}^m\textbf{C}_{k,m-1}\textbf{C}_{k,m-1}^H)$.
%\begin{figure*}
%\small
%\begin{equation}
%\begin{split}
%&\frac{\textbf{w}^H\boldsymbol{\phi}\textbf{w}}{\textbf{w}^H\textbf{Z}\textbf{w}}=
%\frac{\textbf{a}_i^H{\rm diag}\{\textbf{w}_i^H\textbf{C}_{i+1,m-1}\textbf{H}_{i,i+1}
% \textbf{F}_i\}\textbf{C}_{0,i-1}\textbf{C}_{0,i-1}^H{\rm diag}\{\textbf{F}_i^H\textbf{H}_{i,i+1}^H
% \textbf{C}_{i+1,m-1}^H\textbf{w}_i\}\textbf{a}_i}{\textbf{a}_i^H{\rm diag}\{
% \textbf{w}_i^H \textbf{C}_{i+1,m-1}\textbf{H}_{i,i+1}\textbf{F}_i\}(\sum_{k=1}^i\textbf{C}_{k,i-1}\textbf{C}_{k,i-1}^H){\rm diag}\{\textbf{F}_i^H\textbf{H}_{i,i+1}^H\textbf{C}_{i+1,m-1}^H\textbf{w}_i\}\textbf{a}_i
% +\textbf{w}_i^H(\sum_{k=i+1}^m\textbf{C}_{k,m-1}\textbf{C}_{k,m-1}^H)\textbf{w}_i}\\
%&{\rm for}~i=1,~2,...,~m-1
%\end{split}
%\end{equation}
%\end{figure*}

Since the multiplication of any constant value and a eigenvector is
still the eigenvector of the matrix, we can express the receive
filter as
\begin{equation}
\textbf{w}_i =
\frac{\textbf{w}_{opt}}{\sqrt{\textbf{w}_{opt}^H(\sum_{k=i+1}^m\textbf{C}_{k,m-1}\textbf{C}_{k,m-1}^H)\textbf{w}_{opt}}}
\end{equation}
Therefore, we can obtain
\begin{equation}
\textbf{w}_i^H(\sum_{k=i+1}^m\textbf{C}_{k,m-1}\textbf{C}_{k,m-1}^H)\textbf{w}_i = 1 = \frac{N_{i+1}\textbf{a}_i^H\textbf{a}_i}{P_{T,i}}
\end{equation}
By substituting (23) into (21) and using $\textbf{M}_i$ and
$\textbf{N}_i$ given above to simplify the expression of (21), we
obtain
%\begin{figure*}
%\begin{equation}
%\textbf{M}_i = {\rm diag}\{\textbf{w}_i^H\textbf{C}_{i+1,m-1}\textbf{H}_{i,i+1}\textbf{F}_i\}\textbf{C}_{0,i-1}\textbf{C}_{0,i-1}^H{\rm diag}\{\textbf{F}_i^H\textbf{H}_{i,i+1}^H\textbf{C}_{i+1,m-1}^H\textbf{w}_i\},
%\end{equation}
%\end{figure*}
%\begin{figure*}
%\begin{equation}
%\textbf{N}_i = {\rm diag}\{\textbf{w}_i^H\textbf{C}_{i+1,m-1}\textbf{H}_{i,i+1}\textbf{F}_i\}(\sum_{k=1}^i\textbf{C}_{k,i-1}\textbf{C}_{k,i-1}^H){\rm diag}\{\textbf{F}_i^H\textbf{H}_{i,i+1}^H\textbf{C}_{i+1,m-1}^H\textbf{w}_i\}+\frac{N_{i+1}}{P_{T,i}}\textbf{I}
%\end{equation}
%\end{figure*}
\begin{equation}
\frac{\textbf{w}^H\boldsymbol{\phi}\textbf{w}}{\textbf{w}^H\textbf{Z}\textbf{w}}
=
\frac{\textbf{a}_i^H\textbf{M}_i\textbf{a}_i}{\textbf{a}_i^H\textbf{N}_i\textbf{a}_i}~~~~{\rm
for}~i=1,~2,...,~m-1
\end{equation}
We can notice that the expression
$\frac{\textbf{a}^H\textbf{M}_i\textbf{a}}{\textbf{a}_i^H\textbf{N}_i\textbf{a}_i}$
in (26) is the Generalized Rayleigh Quotient, therefore the optimal
solution of our maximization problem can be solved:
$\textbf{a}_{i,opt}$ is any eigenvector corresponding to the
dominant eigenvalue of $\textbf{N}_i^{-1}\textbf{M}_i$, and
satisfying
$\textbf{a}_{i,opt}^H\textbf{a}_{i,opt}=\frac{P_{T,i}}{N_{i+1}}$.
The solutions of $\textbf{w}_{opt}$ and $\textbf{a}_{i,opt}$ depend
on each other. Therefore it is necessary to iterate them with an
initial value of $\textbf{a}_i$ ($i=1,2,...,m-1$) to obtain the
optimum solutions.

\section{Proposed Alternating Maximization Algorithm}

In this section, we devise our proposed alternating maximization
algorithm which computes the linear receive filter and the power
allocation parameters that maximize the sum rate of the WSN. In
particular, we employ two methods to calculate the dominant
eigenvectors. The first one is the QR algorithm \cite{Watkins} which
calculates all the eigenvalues and eigenvectors of a matrix. We can
choose the dominant eigenvector among them. The second one is the
power method \cite{Watkins} which only calculates the dominant
eigenvector of a matrix. Therefore, the computational complexity can
be reduced. Table 1 shows a summary of our proposed algorithm which
will be used for the simulations.

\begin{table}
  \centering
  \caption{Summary of the Proposed Algorithm}\label{}
  \begin{tabular}{l}
  \hline
~~~~~~~~~~~~~~~Initialize the algorithm by setting \\
~~~~~~~~~~~~$ {\textbf{A}_i}=\sqrt{\frac{P_{T,i}}{N_iN_{i+1}}}\textbf{I} $ ~~for $i=1,2,...,m-1$\\
For each iteration:\\
1. Compute $\boldsymbol{\phi}$ and $\textbf{Z}$ in (15) and (16).\\
2. Using QR algorithm or power method to compute the\\~~~ dominate eigenvector of $\textbf{Z}^{-1}\boldsymbol{\phi}$, denoted as $\textbf{w}_{opt}$.\\

3. For $i = 1,2,...,m-1$\\
~~~~a) Compute $\textbf{M}_i$ and $\textbf{N}_i$ in (24) and (25).\\
~~~~b) Using the QR algorithm or power method to compute the\\~~~~~~~~dominate eigenvector of $\textbf{N}_i^{-1}\textbf{M}_i$, denoted as $\textbf{a}_{i}$.\\
~~~~c) To ensure the local power constraint $\textbf{a}_{i,opt}^H\textbf{a}_{i,opt}=\frac{P_{T,i}}{N_{i+1}}$,\\~~~~~~~~compute $\textbf{a}_{i,opt}=\sqrt{\frac{P_{T,i}}{N_{i+1}\textbf{a}_i^H\textbf{a}_i}}\textbf{a}_{i}$.  \\
\hline
\end{tabular}
\end{table}

{\color{red} Tong, we should include a short paragraph here
discussing the required complexity and the convergence issues.}

\section{Simulations}

In this section, we numerically study the sum-rate performance of
our proposed joint MSR design of the receiver and power allocation
methods and compare them with the equal power allocation method
\cite{Luo} which allocates the same power level equally for all
links from the relay nodes. We consider a 3-hop ($m$=3) wireless
sensor network. The number of source nodes ($N_0$), two groups of
relay nodes ($N_1, N_2$) and destination nodes ($N_3$) are 1, 4, 4,
2 respectively. We consider an AF cooperation protocol. The
quasi-static fading channel (block fading channel) is considered in
our simulations whose elements are Rayleigh random variables (with
zero mean and unit variance) and assumed to be invariant during the
transmission of each packet. In our simulations, the channel is
assumed to be known at the destination nodes. For channel estimation
algorithms for WSNs and other low-complexity parameter estimation
algorithms, one can refer to \cite{Wang} and \cite{Lamare1}. During
each phase, the sources transmit the QPSK modulated packets with
1500 symbols. The noise at the destination nodes is modeled as
circularly symmetric complex Gaussian random variables with zero
mean. When perfect (error free) feedback channel between destination
nodes and relay nodes is assumed to transmit the amplification
coefficients, it can be seen from Fig. 3 that our proposed method
can achieve better sum-rate performance than the equal power
allocation method. When using the power method to calculate the
dominant eigenvector, it can get a very similar result to the QR
algorithm. In practice, the feedback channel can not be error free.
In order to study the impact of feedback channel errors on the
performance, we employ the binary symmetric channel (BSC) as the
model for the feedback channel and quantize each complex
amplification coefficient to an 8-bit binary value (4 bits for the
real part, 4 bits for the imaginary part). Vector quantization
methods \cite{Lamare2} can also be employed for increased spectral
efficiency. The error probability (Pe) of BSC is fixed at $10^{-3}$.
The dashed curves in Fig. 3 show the performance degradation
compared with the performance when using a perfect feedback channel.
To show the performance tendency of the BSC for other values of Pe,
we fix the SNR at 10 dB and choose Pe ranging form 0 to $10^{-2}$.
The performance curves are shown in Fig. 4, which illustrates the
sum-rate performance versus Pe of our two proposed methods. It can
be seen that along with the increase in Pe, their performance
becomes worse.

\begin{figure}[!htb]
\centering
\includegraphics[width=3.25in]{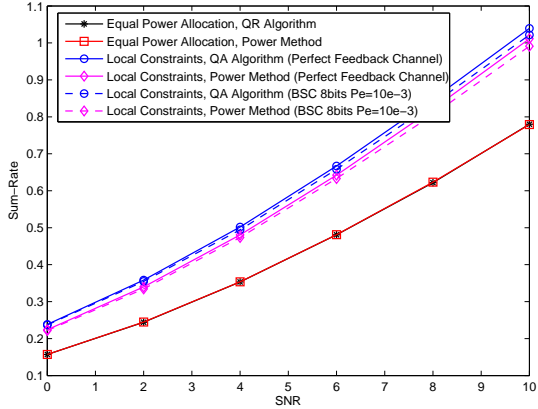}
\vspace{-2em}\caption{\footnotesize Sum-rate performance versus SNR
of our proposed joint maximum sum-rate design of the receiver and
power allocation strategy for a 3-hop WSN, compared with equal power
allocation method.} \vspace{-1em}
\end{figure}
\begin{figure}[!htb]
\centering
\includegraphics[width=3.25in]{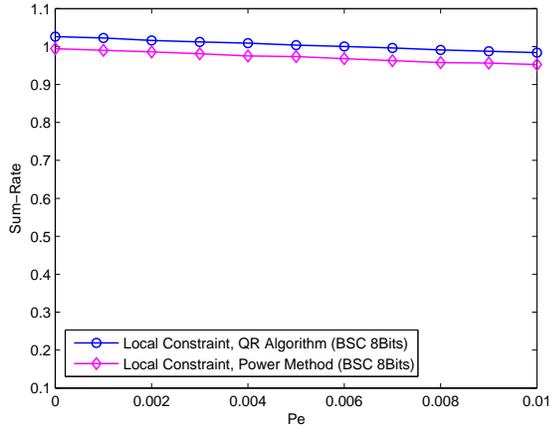}
\vspace{-2em}\caption{\footnotesize Sum-rate performance versus Pe
of our proposed joint strategy when employing BSC as the model for
the feedback channel.} \vspace{-1em}
\end{figure}

\section{Conclusions}

A joint MSR receiver design and power allocation strategy has been
proposed for general multihop WSNs. It has been shown that our
proposed strategy achieves a significantly better performance than
the equal power allocation method. Possible extensions to this work
may include the study of the complexity and the requirement for the
feedback channel.

\section{appendix}
Here, we derive the expression of $\textbf{F}_i$ which is denoted in
Section 2.
\begin{equation}
\textbf{F}_i={\rm diag}\{E(|x_{i,1}|^2),E(|x_{i,2}|^2),...,E(|x_{i,N_i}|^2)\}^{-\frac{1}{2}}
\end{equation}
where
\begin{equation}
E(|x_{i,j}|^2 = \left\{ \begin{array}{ll}
\sigma_s^2|\textbf{h}_{s,j}|^2+\sigma_n^2, ~~~~~~~~~~~~~~~~~~~~~~~~~~~~~~~~~~~~~~\textrm{for $i=1$,}\\\textbf{h}_{i-1,j}\textbf{A}_{i-1}E(\textbf{y}_{i-1}\textbf{y}_{i-1}^H)\textbf{A}_{i-1}^H\textbf{h}_{i-1,j}^H+\sigma_n^2,\\
~~~~~~~~~~~~~~~~~~~~~~~~~~~~~~~~~~~~~~~~~~~~~~~~~~\textrm{for $i=2,3,...,m$}.\\
\end{array}\right.
\end{equation}

\begin{equation}
E(\textbf{y}_{i}\textbf{y}_{i}^H) = \left\{ \begin{array}{ll}
\textbf{F}_i(\sigma_s^2\textbf{H}_s\textbf{H}_s^H+\sigma_n^2\textbf{I})\textbf{F}_i^H, ~~~~~~~~~~~~~~~~~~~~~~~~~ \textrm{for $i=1$,}\\\textbf{F}_i[\textbf{H}_{i-1,i}\textbf{A}_{i-1}E(\textbf{y}_{i-1}\textbf{y}_{i-1}^H)\textbf{A}_{i-1}^H\textbf{H}_{i-1,i}^H
+\sigma_n^2\textbf{I}]\textbf{F}_i^H\\
~~~~~~~~~~~~~~~~~~~~~~~~~~~~~~~~~~~~~~~~~~~~~~~~~~~~\textrm{for $i=2,3,...,m$}.\\
\end{array}\right.
\end{equation}

\bibliographystyle{IEEEbib}

\end{document}